\documentclass[12pt,a4paper]{article}

\usepackage[fleqn]{amsmath}
\usepackage{alltt}
\usepackage{amssymb}
\newtheorem{definition}{Definition}
\newtheorem{theorem}{Theorem}
\newtheorem{lemma}[theorem]{Lemma}

\newtheorem{corollary}[theorem]{Corollary}

\newcommand{\done}{$\Box$\bigskip}

\def\x{$\hfill\rlap{$\sqcup$}\sqcap$\bigskip}

\newcommand{\F}{\mathbb F}

\newcommand{\tfr}[1]{{\widehat{#1}}}
\def\Tr{{\rm Tr}}

\title{Fourier Spectra of Binomial APN Functions}

\author{Carl Bracken\thanks{School of Mathematical Sciences,
University College Dublin,
Ireland.
({\tt carlbracken@yahoo.com})
Research supported by
Irish Research Council for Science, Engineering and Technology
Postdoctoral Fellowship.}
\and
Eimear Byrne\thanks{School of Mathematical Sciences,
University College Dublin,
Ireland.
({\tt ebyrne@ucd.ie})
Research supported by the Claude
Shannon Institute, Science Foundation Ireland Grant 06/MI/006.}
\and
Nadya Markin\thanks{School of Mathematical Sciences,
University College Dublin,
Ireland.
({\tt nadyaomarkin@gmail.com})
Postdoctoral Fellow supported by the Claude
Shannon Institute, Science Foundation Ireland Grant 06/MI/006.}
\and
Gary McGuire\thanks{School of Mathematical Sciences,
University College Dublin,
Ireland.
({\tt gary.mcguire@ucd.ie})
Research supported by the Claude
Shannon Institute, Science Foundation Ireland Grant 06/MI/006.}}

\begin{document}
 
\maketitle

\bigskip
\begin{abstract}
\noindent 
In this paper we compute the Fourier spectra of some recently discovered binomial APN functions.
One consequence of this is the determination of the nonlinearity of the functions,
which measures their resistance to linear cryptanalysis.
Another consequence is 
that certain error-correcting codes related to these functions 
have the same weight distribution as the 2-error-correcting BCH code. 
Furthermore, for field extensions of $\mathbb{F}_2$ of odd degree, our results provide an 
alternative proof of the APN property of the functions.

\end{abstract}

\pagestyle{myheadings}
\thispagestyle{plain}

\bigskip

\section{Introduction}
Highly nonlinear functions on finite fields are interesting from the point of view of cryptography as they provide optimum resistance to linear and differential attacks. A function that has the APN (resp. AB) property, as defined below, has optimal resistance to a differential (resp. linear) attack. 
For more on relations between linear and differential cryptanalysis, see \cite{CV}.

Highly nonlinear functions are also of interest from the point of view of coding theory. 
The weight distribution of a certain error-correcting code is equivalent to the 
Fourier spectrum (including multiplicities) of $f$.
The code having three particular weights is equivalent to the AB property, when $n$ is odd.
The minimum distance of the dual code being 5 is equivalent to the APN property holding for $f$.
We give more details on the connections to coding theory in Section 2.

For the rest of the paper, let $L = GF(2^n)$ 
and let $L^*$ denote the set of non-zero elements of $L$. 
Let $\Tr: L \rightarrow GF(2)$ denote the trace map from $L$ to $GF(2)$. 

\begin{definition}
A function $f: L \rightarrow L$ is said to be {\emph{\bf{ almost perfect nonlinear (APN)}}} 
if for any $a\in L, b \in L^*$,  we have
$$ |\{x \in L : f(x+a)-f(x) = b \}| \leq 2. $$
\end{definition} 

\begin{definition}
Given a function $f:L \rightarrow L$, the {\emph {\bf{Fourier transform}}} of $f$ is the function 
$\tfr f: L \times L^* \rightarrow \mathbb Z$ given by      
$$\tfr f(a,b) = \sum_{x \in L}(-1)^{\Tr(ax+bf(x))}.$$
\label{WT}
\end{definition}

The {\emph{Fourier spectrum}} of $f$ is the set of integers
$$\Lambda_f= \{\tfr f(a,b) : a, b \in L, b \neq 0 \}.$$ 

\newpage
\noindent
The nonlinearity of a function $f$ on a field $L=GF(2^n)$ is defined as
$$NL(f):= 2^{n-1} - \frac{1}{2}\max_{x \in \Lambda_f} \ |x| .$$
The nonlinearity of a function measures its distance 
to the set of all affine maps on $L$.
We thus call a function {\emph{maximally nonlinear}} if its nonlinearity is as large as possible.
If $n$ is odd, its nonlinearity is upper-bounded by $2^{n-1}-2^{\frac{n-1}{2}}$, while for $n$ even 
an upper bound is  $2^{n-1}-2^{\frac{n}{2}-1}$.
For odd $n$, we say that a function $f : L\longrightarrow L$ is \emph{almost bent} (AB) 
when its Fourier spectrum is $\{0, \pm 2^{\frac{n+1}{2}}\} $, 
in which case it is clear from the upper bound that $f$ is maximally nonlinear.
We have the following connection (for odd $n$)
between the AB and APN property: every AB
function on $L$ is also APN \cite{CV}, and, conversely, 
if $f$ is quadratic and APN, then $f$ is AB \cite{CCZ}. 
In particular, quadratic APN functions have optimal resistance to both linear and differential attacks.
 On the other hand, there appears to be no relation between the nonlinearity and the APN property of a function when $n$ is even. 
The reader is referred to \cite{C} for a comprehensive survey on APN and AB functions.

Recently, the first non-monomial families of APN functions have been discovered.
Below we list the families of quadratic functions known at the time of writing.
We remark that, in a sense to be qualified in the next section, these families are all
pairwise inequivalent.

\bigskip

\begin{enumerate}
\item
$$f(x)= x^{2^s+1}+ \alpha x^{2^{ik}+2^{mk+s}},$$
where $n =3k$, $(k,3)=(s,3k)=1$, $k \geq 3$, $i \equiv sk \mod 3$, $m \equiv -i \mod 3$, $\alpha = t^{2^k-1}$ and $t$ is primitive (see 
 Budaghyan,  Carlet,   Felke,  Leander \cite{BCFL}).

\bigskip

\item

$$f(x)= x^{2^s+1}+ \alpha x^{2^{ik}+2^{mk+s}},$$ where 
$n =4k$, $(k,2)=(s,2k)=1$, $k \geq 3$, $i \equiv sk \mod 4$, $m= 4-i$, $\alpha = t^{2^k-1}$ and $t$ is primitive (see Budaghyan,  Carlet,  Leander \cite{BCL2}).
This family generalizes an example found for $n=12$ by
 Edel, Kyureghyan, Pott \cite{EKP}.

\bigskip

\item
$$f(x) = \alpha x^{2^{s}+1}+{\alpha}^{2^k}x^{2^{k+s}+2^k}+\beta
x^{2^k+1}+\sum_{i=1}^{k-1} {\gamma}_i x^{2^{k+i}+2^i},$$
where $n=2k$,
$\alpha$ and $\beta$ are primitive elements of $GF(2^{n})$, and
${\gamma}_i \in GF(2^k)$ for each $i$, and $(k,s)=1$, $k$ is odd,
$s$ is odd (see  Bracken,  Byrne,  Markin,  McGuire \cite{BBMMcG}).

\bigskip

\item
$$f(x) = x^3 + Tr(x^9),$$
over $GF(2^n),$ any $n$ (see Budaghyan,  Carlet,  Leander  \cite{B+}).

\bigskip

\item
$$f(x)= ux^{2^{-k}+2^{k+s}}+u^{2^{k}}x^{2^{s}+1}+vx^{2^{k+s}+2^{s}},$$
where $n=3k$,
$u$ is primitive, $v \in GF(2^k), (s,3k)=1,\  (3,k)=1$ and 3 divides $k+s$
(see Bracken,  Byrne,  Markin,  McGuire \cite{BBMMcG}). 
\end{enumerate}

In this paper we calculate the Fourier spectra of families (1) and (2).
The determination of the Fourier spectra of families (3) and (4)
has been given in \cite{BBMMcG2} and \cite{BBMMcG3}, respectively, 
using other methods. The Fourier spectrum for family (5) has not yet been found, and
is an open problem.
We will show here that the Fourier spectra of the functions (1) and (2)
are 5-valued for fields of even degree and 3-valued for fields of odd degree. 
In this sense they resemble the Gold functions $x^{2^d+1}$, $(d,n)=1$.
For fields of odd degree, our result provides another proof of the APN property.
This does not hold for fields of even degree;
as we stated earlier, 
there appears to be no relation between the Fourier spectrum and the APN property
for fields of even degree.
Thus, the fact that $f$ has a $5$-valued Fourier 
spectrum for fields of even degree does not follow from the fact that $f$ is a quadratic APN function. 
Indeed, there is one example known (due to Dillon \cite{Dillon})
of a quadratic APN function on a field of even degree whose
Fourier spectrum is more than 5-valued;
if $u$ is primitive in $GF(2^6)$ then
$$g(x)=x^3+u^{11}x^5+u^{13}x^9+x^{17}+u^{11}x^{33}+x^{48}$$
is a quadratic APN function on $GF(2^6)$ whose Fourier transform takes seven distinct values.

The layout of this paper is as follows.
In Section 2 we review the connections between APN functions,
nonlinearity, and coding theory.
Section 3 gives the proof of the Fourier spectrum for family (1),
and Section 4 gives the proof for family (2).
In Section 5 we simply state for completeness
 the results from other papers on families (3) and (4),
 and Section 6 has some open problems for further work.

\section{Preliminaries on Coding Theory}\label{coding}

Fix a basis of $L$ over $\F_2$. For each element $x \in L$
we write ${\bf x}= (x_1,...,x_n)$ to denote the vector of coefficients of $x$
with respect to this basis. Given a map $f : L\longrightarrow L$, 
we write $f({\bf x})$ to denote
the representation of $f(x) \in L$ as a vector in $\F_{2}^n$, and  
we consider the $2n \times (2^n-1)$ binary matrix
$$A_f=\left[ \begin{array}{ccc}
                       \cdots             &{\bf x}& \cdots \\
                         \cdots             &f({\bf x})& \cdots 
                     \end{array} \right] $$
where the columns are ordered
with respect to some ordering of the nonzero elements of $L$

The function $f$ is APN if and only if the binary error-correcting code 
of length $2^n-1$ with $A_f$ as 
parity check matrix
has minimum distance 5. 
This is because codewords of weight 4 correspond to solutions of
\begin{eqnarray*}
a+b+c+d&=&0\\
f(a)+f(b)+f(c)+f(d)&=&0
\end{eqnarray*}
and this system has no nontrivial solutions if and only if $f$ is APN.
We refer the reader to \cite{CCD} for more on the connection between 
coding theory and APN functions. The dual code has $A_f$ as a generator matrix. 
The weights $w$ in this code correspond to values $V$ in the Fourier spectrum
of $f$ via $V=n-2w$.
Thus, when we compute the Fourier spectrum of an APN function,
as we do in this paper, we are computing the weights occurring in the code.

Suppose $f$ is APN.
Let $C_f$ denote the code with generator matrix $A_f$.
Let $a_w$ denote the number of times the weight $w$ occurs in $C_f$.
Let $b_j$ denote the number of codewords of weight $j$ in $C_f^\perp$.
If there are five or fewer weights in $C_f$, the MacWilliams (or Pless)  identities 
yield five independent equations $b_0=1$, $b_1=b_2=b_3=b_4=0$,
for the unknowns $a_V$, which can be solved uniquely.
Thus the distribution of values is determined for an APN function
whenever there are five or fewer values in its Fourier spectrum. 
In particular, if  $\Lambda_f \subseteq \{0,\pm 2^{\frac{n}{2}},\pm 2^{\frac{n+2}{2}}\}$
for even $n$, or 
$\Lambda_f \subseteq \{0,\pm 2^{\frac{n+1}{2}}\}$ for odd $n$,
then the distribution is completely determined.
This is indeed the case for the functions studied in this paper.
Solving for the distribution in this case must yield the same values and
distribution as the double-error-correcting BCH code, which corresponds
to the APN function $x^3$.
This function has
 $\Lambda_f = \{0,\pm 2^{\frac{n}{2}},\pm 2^{\frac{n+2}{2}}\}$
for even $n$, and 
$\Lambda_f = \{0,\pm 2^{\frac{n+1}{2}}\}$ for odd $n$,


Consider the extended code of $C_f^\perp$, which has parity check matrix
$$P_f=\left[ \begin{array}{cccc}
                         1     \cdots             &1&     1    &1\\
                       \cdots             &x& \cdots &0\\
                         \cdots             &f(x)& \cdots&0 
                     \end{array} \right] .$$
Two functions $f$ and $g$ are said to be CCZ equivalent if and only if
the codes with parity check matrices $P_f$ and $P_g$ are equivalent (as binary codes).
This is not the original definition of CCZ equivalence, but it is
an equivalent definition, as was shown in \cite{BBMMcG}.

The new APN functions presented in the introduction
are known to be pairwise CCZ-inequivalent.
One consequence of the results in this paper is that further
invariants (beyond the code weight distribution) are needed 
to show that families (1)-(4) are inequivalent.

\section{Family (1), Binomials over $GF(2^{3k})$}
We will make good use of the following standard result from Galois theory, 
which allows us to bound the number of solutions of a linearized polynomial. 
We include a proof for the convenience of the reader.

\begin{lemma}
Let $F$ be a field and let $K,H$ be finite Galois extensions of $F$ of degrees $n$ and 
$s$ respectively, whose intersection is $F$. Let $M = KH$ be the compositum of $K$ and $H$. Let $k_1, \ldots, k_t$ be $F$-linearly independent elements of $K$. Then $k_1, \ldots, k_t$ are $H$-linearly independent when regarded as elements of $M$. 
\label{lin}\end{lemma}

\noindent{\bf Proof:}\\
Since $K$ and $H$ are Galois extensions of $F$ and $K \bigcap H = F$, we have $[M:H] = [K:F] = n$.  Let $\{k_1, \ldots, k_n\}$ be an $F$-basis of $K$ as a vector space over $F$, $\{h_1, \ldots, h_s\}$ an $F$-basis of $H$ as a vector space over $F$. Then the set $\{k_i\cdot h_j \mid 1 \leq i \leq n, 1\leq j \leq s\}$ generates $M$ as a vector space over $F$. It is clear that the set $\{k_1 \cdot h_1, \ldots, k_n \cdot h_1\}$ generates $M$ as a vector space over the field $H$. Without loss of generality we can assume that $h_1 = 1$. Since $[M:H]=n$, we conclude that $\{k_1, \ldots, k_n\}$ is indeed a basis of $M$ over $H$. 

Let $\{k_1, \ldots, k_t\}$ be a set of $F$-linearly independent elements of $K$. We can extend this set to a basis $\{k_1, \ldots, k_t, \ldots, k_n\}$. Since this set forms an  $H$-basis of $M$, its subset $\{k_1, \ldots, k_t\}$ is a fortiori linearly independent over $H$.
\done

Note that for Galois extensions $K,H$ in the lemma above, $(s,n)=1$ implies that $K\bigcap H = F$ and in the case when the fields $K,H,F$ are finite, we have $(s,n)=1 $ if and only if $K \bigcap H = F$.

\begin{corollary}
Let $s$ be an integer satisfying $(s,n)=1$ and let 
$f(x) =\displaystyle{ \sum_{i=0}^d r_i x^{2^{si}}}$ be a polynomial in $ L[x]$. Then $f(x)$ has at most $2^d$ zeroes in $L$. 
\label{rsol}
\end{corollary}

\medskip

Let $V$ denote the set of zeroes of $f(x)$ in $L$. We may assume that $V \neq \{0\}$.
Since $f(x)$ is a linearized polynomial, $V$ is a vector space over $GF(2)$ of finite dimension $v$ for some positive integer $v$. Let $V' \subset GF(2^{sn})$ denote the vector space generated by the elements of $V$ over the field $GF(2^{s})$. Since $(s,n)=1$, we have $
L \bigcap GF(2^s) = GF(2)$ and by Lemma \ref{lin}, $V'$ is a $v$-dimensional vector space over $GF(2^s)$. Furthermore, for all $c \in GF(2^s)$ and $w \in GF(2^{sn})$ we have $f(cw)=cf(w)$. Therefore all the elements of $V'$ are also zeroes of $f(x)$. Since the dimension of $V$ over $GF(2)$ is $v$, the size of $V'$ is $2^{sv}$ and it follows that there are at least $2^{sv}$ zeroes ot $f(x)$ in $GF(2^{sn})$. On the other hand, polynomial of degree $2^{ds}$ can have at most $2^{ds}$ solutions. We conclude that $v \leq d$.  
\done

\begin{theorem}
Let $$f(x)= x^{2^s+1}+ \alpha x^{2^{ik}+2^{mk+s}},$$

where $n =3k$, $(k,3)=(s,3k)=1$, $k \geq 3$, $i \equiv sk \mod 3$, $m \equiv -i \mod 3$, $\alpha = t^{2^k-1}$ and $t$ is primitive in $L$. 
 
The Fourier spectrum of $f(x)$ is $\{0, \pm 2^{\frac{n+1}{2}}\} $ when $n$ is odd and $\{0, \pm 2^{\frac{n}{2}}, \pm 2^{\frac{n+2}{2}} \}$
when $n$ is even.

\end{theorem}
Proof:\\
\noindent
By the restrictions on $i,s,k$, there are two possibilities for our function $f(x)$:

$$f_1(x)= x^{2^s+1}+ \alpha x^{2^{-k}+2^{k+s}} \hspace{1 cm} sk \equiv -1 \mod 3$$  and
$$f_2(x)= x^{2^s+1}+ \alpha x^{2^{k}+2^{-k+s}} \hspace{1 cm} sk \equiv 1 \mod 3.$$  

Let us consider the first case, when $f = f_1$. By definition, the Fourier spectrum of $f$ is 

$$f^W(a,b) = \sum_u(-1)^{Tr(ax+ b f(x))}. $$
Squaring gives
$$ f^W(a,b)^2 = \sum_{x \in L} \sum_{u \in L}(-1)^{\Tr(ax+bf(x)+a(x+u)+bf(x+u))}.$$
This becomes
$$f^W(a,b)^2 = \sum_u{(-1)^{Tr(au+bu^{2^s+1}+ b \alpha u^{2^{-k}+2^{k+s}})}}\sum_x{(-1)^{Tr(xL_b(u))}},$$
where $L_b(u): = bu^{2^s}+(bu)^{2^{-s}}+(b \alpha)^{2^k}u^{2^{-k+s}}+ (b\alpha)^{2^{-k-s}}u^{2^{k-s}}.$

Using the fact that $ \sum_x(-1)^{Tr(cx)}$ is $0$ when $c \neq 0$ and $2^n$ otherwise, we obtain 

$$f^W(a,b)^2 = 2^n \sum_{u \in K} (-1)^{Tr(au+bu^{2^s+1}+b \alpha u^{2^{-k}+2^{k+s}})},$$

where $K$ denotes the kernel of $L_b(u)$. If the size of the kernel is at most 4, 
then clearly
\[
0\leq \sum_{u \in K} (-1)^{Tr(au+bu^{2^s+1}+b \alpha u^{2^{-k}+2^{k+s}})} \leq 4.
\]
Since  $f^W(a,b)$ is an integer, this sum can only be 0, 2, or 4 if $n$ is even, and 1 or 3 if $n$ is odd.
 The set of permissible values of $f^W(a,b)$  is then


$$ f^W(a,b) \in
\begin{cases}
\{0, \pm 2^{\frac{n+1}{2}} \} & 2 \nmid n \\
\{ 0, \pm 2^{\frac{n}{2}}, \pm 2^{\frac{n+2}{2}} \} & 2 \mid n.
\end{cases}
$$

We must now demonstrate that $|K| \leq 4$, which is sufficient to complete the proof.

Note that since $\alpha$ is a $(2^k-1)$-th power, we have $\alpha^{2^{2k} + 2^k+1}=1$. 
Now suppose that $L_b(u)=0$. Then we have the following equations:

$$(b  \alpha)^{-2^k}L_b(u) 
+ b^{1-2^k-2^{-k}}\alpha L_b(u)^{2^k}+b^{-2^{-k}} L_b(u)^{2^{2k}}=0,$$

$$b^{-2^{-s}}L_b(u) + 
b^{2^{-k-s}-2^{k-s}-2^{-s}}\alpha^{2^{-k-s}}L_b(u)^{2^k} + 
b^{-2^{k-s}}\alpha^{-2^{k-s}}L_b(u)^{2^{-k}}=0.$$
Substituting the definition of $L_b(u)$ into equations above and gathering the terms gives

\begin{equation}\label{3keq1}
$$ \ \ \ \ \ \ \ \ \ \ \ \ \ \ \ \ \ \ \ \ c_1 u^{2^{-s}}+ c_2 u^{2^{k-s}} + c_3 u^{2^{-k-s}} =0, $$
\end{equation}

\begin{equation}\label{3keq2}
$$ \ \ \ \ \ \ \ \ \ \ \ \ \ \ \ \ \ \ \ \ d_1 u^{2^s} + d_2 u^{2^{k+s}}+  d_3 u^{2^{-k+s}} = 0,$$
\end{equation}
\noindent
where\\
\noindent
$ c_1 = (b^{2^{-s}-2^{k}} \alpha^{-2^k} + b^{2^{k-s}-2^{-k}} \alpha^{2^{k-s}}), \\
c_2 = ((b \alpha)^{2^{-k-s}-2^{k}} +b^{2^{k-s}+1-2^{-k}-2^k} \alpha), \\
c_3 = (b^{2^{-s}+1 -2^k - 2^{-k}} \alpha^{2^{-s}+1} + b^{2^{-k-s}-2^{-k}}),$ \\
$d_1 =(b^{1-2^{-s}}+ b^{2^{-k-s}+2^{-k}-2^{-s}-2^{k-s}} \alpha^{2^{-k-s}+2^{-k}}), \\
d_2 = (b^{2^{-k-s}+2^k-2^{-s}-2^{k-s}} \alpha^{2^{-k-s}}+b^{1-2^{k-s}} \alpha^{2^{-k-s}+2^{-s}+1}),\\
d_3 = (b^{2^k-2^{-s}} \alpha^{2^k} + b^{2^{-k}-2^{k-s}} \alpha^{2^{-k-s}+2^{-s}}).$

First we demonstrate that the coefficients $c_i,d_j$ in Equations (\ref{3keq1}) and (\ref{3keq2}) do not vanish. Suppose that $c_1=0$. 
We then have 
$$\alpha^{2^{k-s}+2^k} = b^{-2^{k-s}+2^{-k}+2^{-s}-2^{k}}$$ and taking $2^{-k}$-th power of both sides yields
$$\alpha^{2^{-s}+1} = b^{(2^{k+s}-1)(2^{-s}-2^{-k-s})}.$$ 
Let $\alpha = t^{2^k-1}$, where $t$ is primitive in $GF(2^{3k})$. Substituting $t$ into the previous equation and some rearrangement gives
$$t^{2^{k-s}-1}=t^{2^{-s}(1-2^{k+s})} b^{(2^{k+s}-1)(2^{-s}-2^{-k-s})}.$$ 
The multiplicative order of 2 modulo 7 is equal to 3, therefore for any $r$ we have 
$7$ divides $2^r-1$ if and only if  $r$ is divisible by $3$. 
Since $3 \nmid k-s$, we conclude that $7 \nmid 2^{k-s}-1$, therefore the left hand side is not a seventh power, while the right hand side is.  
We conclude that the coefficient of $u^{2^{-s}}$ in Equation (\ref{3keq1}) is not $0$ and use the same type of argument to conclude that all the coefficients in Equation (\ref{3keq1}) are non-zero. 
A similar argument holds for Equation (\ref{3keq2}).

We will next combine Equation (\ref{3keq1}) and Equation (\ref{3keq2}) to obtain an equation of the form
$$ Au + Bu^{2^k} = 0.$$ 
Raise Equation (\ref{3keq1}) to the power of $2^s$, Equation (\ref{3keq2}) to the power of $2^{-s}$ and combine the two expressions, cancelling the terms in $u^{2^{-k}}$ to obtain

\begin{equation} $$\ \ \ \ \ \ \ \ \ \ \ \ \ \ \ \ \ \ \ \ \ \ \ \ \ \ \ \ \ \ \ \ Au + Bu^{2^k} = 0,$$  \label{3keq3} \end{equation}
where $A = (\frac{c_1}{c_3})^{2^s} + (\frac{d_1}{d_3})^{2^{-s}}$ and $B = (\frac{c_2}{c_3})^{2^s}+(\frac{d_2}{d_3})^{2^{-s}}.$

For now assume that both $A,B$ are non-zero. We obtain the following equalities by applying the appropriate powers of the Frobenius automorphism to Equation (\ref{3keq3}):

$$ u^{2^{-k+s}} = A^{-2^{-k+s}} B^{2^{-k+s}}u^{2^s}, $$

$$u^{2^{k-s}}=B^{-2^{-s}}A^{2^{-s}} u^{2^{-s}}.$$

Substituting the two identities above to our expression for $L_b(u)=0$ gives 
\begin{equation}\label{3keq4}
$$(b+(b \alpha)^{2^k} A^{-2^{-k-s}}B^{2^{-k-s}})u^{2^s} +  (b^{2^{-s}}+(b \alpha)^{2^{-k-s}} B^{-2^{-s}}A^{2^{-s}})u^{2^{-s}} =0.$$
\end{equation}
Raising this equation to the power of $2^s$ gives a polynomial of degree $2^{2s}$ 
which is $GF(2^s)$-linear. By Corollary \ref{rsol}, the dimension of the kernel of this polynomial over $GF(2)$ is at most $2$, unless the lefthand side of Equation  (\ref{3keq4}) is identically $0$. 
It therefore remains to show that the polynomial in Equation (\ref{3keq4}) is not identically $0$. Assuming that both coefficients are zero, we get

$$ A b^{2{k-s}}+ (b \alpha)^{2^{-k-s}}B = 0,$$
$$Bb + (b \alpha)^{2^{-k}}A = 0.$$

We combine the equations above to obtain 

$$Bb + (b \alpha)^{2^{-k}}b^{2^{-k-s}-2^{k-s}} \alpha^{2^{-k-s}}B =0.$$

So we have $b^{1-2^{-k}+2^{k-s}-2^{-k-s}}=\alpha^{2^{-k-s}+2^{-k}}$. Substituting $\alpha$ with $t^{2^k-1}$, rearranging and factoring the powers gives

$$b^{ (2^{k+s}-1)(1-2^{-k})}t^{1-2^{k+s}} = t^{2^s(2^{k-s}-1)}.$$

Here we observe that only the left hand side of the above equation is a seventh power, thus obtaining the desired contradiction. We conclude that the size of the kernel $K$ is less than $4$. This finishes the argument.

It finally remains to show that the coefficients $A, B$ are non-zero. 
Setting $A$ to $0$ gives rise to the equation

$\alpha^{2^{k-2s}+2^{k+s}} = $
$$\left(\frac
{b^{1-2^{-k+s}}+   (b\alpha)^{2^k+2^{k+s}}}
{b^{2^{-s}+2^{k-2s}}+b^{2^{-k}+2^{-k-s}}\alpha^{2^{-k-2s}+2^{-k-s}}}\right)
\left(\frac
{(b\alpha)^{2^{k-s}+2^{k-2s}} + b^{2^{-k-s}-2^{k-2s}} }
{(b \alpha)^{2^s+1} +b^{2^{-k}+2^{k+s}}}\right).$$ 

Substituting $\alpha$ with $t^{2^k-1}$ and rearranging gives the equation 

$$ t^{2^{k-2s+1}(2^k-1)} = t^{2^{k-2s}-2^{-k-2s}(2^{3s}-1)} R^{2^{2k+2s}-1}T^{1-2^{2k+2s}},$$

where $$R = b^{2^{-s}+2^{k-2s}}+b^{2^{-k}+2^{-k-s}}\alpha^{2^{-k-2s}+2^{-k-s}}$$ and 
$$T = ((b \alpha)^{2^{k-s}+2^{k-2s}} +b^{2^{-k-s}-2^{-k-2s}}).$$
Reducing the powers of $2$ modulo 3 shows that the right hand side of the equation above is a seventh power, while the left hand side is not. We conclude that $A \neq 0$.

Suppose $B=0$, then the only solution of Equation (\ref{3keq3}) is $u=0$. We can therefore assume that both $A$ and $B$ are non-zero. 

This completes the proof of the theorem for the case when $f = f_1$. When $f=f_2$ a similar proof applies. We interchange $k$ and $-k$ in all equations and use the fact that in this case $3$ divides $k-s$.

\section{Family (2), Binomials over $GF(2^{4k})$}

We now compute the Fourier spectrum for family (2).

\begin{theorem}
Let $L = GF(2^n)$ and $f(x)= x^{2^s+1}+ \alpha x^{2^{ik}+2^{mk+s}}$, where
$n =4k$, $(k,2)=(s,2k)=1$, $k \geq 3$, $i \equiv sk \mod 4$, $m= 4-i$, $\alpha = t^{2^k-1}$ and $t$ is primitive. Then $f$ has Fourier spectrum $\{ 0, \pm 2^{n/2},  \pm 2^{\frac{n+2}{2}} \}$.
\end{theorem}

Proof:
Since $s,k$ are chosen to be odd, $sk \equiv \pm 1 \mod 4$. Therefore there are two possibilities for our function $f(x)$:

$$f_1(x)= x^{2^s+1}+ \alpha x^{2^{-k}+2^{k+s}} \hspace{1 cm} sk \equiv -1 \mod 4$$  and
$$f_2(x)= x^{2^s+1}+ \alpha x^{2^{k}+2^{-k+s}} \hspace{1 cm} sk \equiv 1 \mod 4.$$  

Let us consider the first case, when $f = f_1$. As discussed in the proof of the previous theorem, since $f$ is APN, it suffices to demonstrate that the equation 

$$L_b(u) = bu^{2^s}+(bu)^{2^{-s}}+(b \alpha)^{2^{mk}}u^{2^{2k+s}}+ (b\alpha)^{2^{ik-s}}u^{2^{2k-s}} =0.$$
has at most four solutions for all non-zero $b$ in $L$. 

All the solutions of $L_b(u)=0$ are also solutions of the equation 
$$b^{-2^{2k}} L_b(u)^{2^{2k}} + (b \alpha)^{-2^k}L_b(u) = 0.$$

This gives 
$$(b^{2^{2k}+1}+ (b \alpha)^{2^k+2^{-k}})u^{2^s} +  
(b^{2^{2k}+2^{-s}} + (b \alpha)^{2^k+2^{k-s}})u^{2^{-s}} +$$
\begin{equation} \label{4k1}
\ \ \ \ \ \ \ \ \ \ (b^{{2^k}+2^{2k-s}}\alpha^{2^{2k-s}}+ b^{2^{2k}+2^{-k-s}} \alpha^{2^{-k-s}})u^{2^{2k-s}} = 0.
\end{equation}

We also compute $b^{-2^{-s +2k}}L_b(u)^{2^{2k}}+ (b \alpha)^{-2^{-k-s}} L_b(u) = 0$ to obtain
$$ (b^{2^{2k-s}}+(b \alpha)^{2^{-k-s}+2^{-k}})u^{2^s}+
(b^{2^{2k-s}+2^{-s}}+(b \alpha)^{2^{k-s}+2^{-k-s}})u^{2^{-s}}+$$
\begin{equation} \label{4k2}
\ \ \ \ \ \ \ \ \ \ \ (b^{2^{2k-s}+2^k} \alpha^{2^k} +b^{2^{2k}+2^{-k-s}}\alpha^{2^{-k-s}})u^{2^{2k+s}}=0.
\end{equation}
Writing Equation (\ref{4k2}) as

\begin{equation} \ \ \ \ \ \ \ \ \ \ \ \ \ \ \ \ \ \ \ \  cu^{2^s}+d u^{2^{-s}}+e u^{2^{2k+s}}=0 \label{4k3}\end{equation} we see that Equation (\ref{4k1}) becomes

\begin{equation}  \ \ \ \ \ \ \ \ \ \ \ \ \ \ \ \ \ \ \ \  d^{2^s} u^{2^s}+ c^{2^{2k}} u^{2^{-s}}+ e u^{2^{2k-s}}=0. \label{4k4} \end{equation}

We combine Equations (\ref{4k3}) and (\ref{4k4}) to cancel the third term from each expression. This yields the following equation
\begin{equation}\label{4kG}
\ \  G(u):=(e^{2^s}c^{2^{-s}}+e^{2^{-s}}c^{2^{2k+s}})u + 
e^{2^s}d^{2^{-s}} u^{2^{-2s}} + e^{2^{-s}}d^{2^{2s}}u^{2^{2s}} = 0.
\end{equation}

Now for some non-zero $v$ in the kernel of $G(u)$, we consider the equation
\begin{equation}\ \ \ \ \ G_v(u):= uG(u)+vG(v) + (u+v)G(u+v)=0. \label{4kGv}\end{equation}

Substituting gives

\begin{equation}
\ \ \ \ e^{2^s}d^{2^{-s}}(u^{2^{-2s}}v+ v^{2^{-2s}}u)+ e^{2^{-s}}d^{2^{2s}}(u^{2^{2s}}v + v^{2^{2s}}u)=0. \label{main} \end{equation}
Note that $ker(G(u))$ is contained in $ker (G_v(u))$. 

We now show that $L_b(u)=0$ has at most four solutions. 
This will be done in five steps, which complete the proof.


\begin{enumerate}

\item[(i)]{We show that $d \neq 0$ implies that $d^{2^s-1}$ is not a cube.} 

Recall that $d = b^{2^{2k-s}+2^{-s}}+b^{2^{k-s}+2^{-k-s}}t^{2^{2k-s}+2^{-s}-2^{k-s}-2^{-k-s}}.$ This implies that $$d^{2^s-1} = t^{-2^{-s-k}(2^{2k}+1)(2^s-1)}A^{2^s-1},$$ where $A = b^{2^{2k-s}+2^{-s}}t^{2^{k-s}+2^{-k-s}}+b^{2^{k-s}+2^{-k-s}}t^{2^{2k-s}+2^{-s}}$. As $A = A^{2^k}$, we have $A \in GF(2^k)$. Furthermore, as $k$ is odd, all elements of $GF(2^k)$ are cubes. We conclude that $A^{2^s-1}$ is a cube. Now if $d^{2^s-1}$ is a cube, then so is $t^{(2^{2k}+1)(2^s-1)}$. But this is impossible as $(2^{2k}+1)(2^s-1)$ is not divisible by $3$ and $t$ is primitive.

\item[(ii)]{We show that if $c,d,e \neq 0$ and $d^{2^s-1}$ is not a cube then 
(\ref{main}) has at most four solutions}. 

Assume that the coefficients $e,c,d$ are non-zero and that $d^{2^s-1}$ is not a cube. 
Now $u^{2^{2s}}v+v^{2^{2s}}u=0$ if and only if $uv^{-1} \in GF(4)$. Therefore we have exactly four solutions in $u$, namely $u=vw$ for each $w \in GF(4)$. 
If, on the other hand, $u^{2^{2s}}v+v^{2^{2s}}u \neq 0$, we can rearrange (\ref{main}) to obtain 

$$d^{2^s-1} = (u^{2^{-2s}}v + v^{2^{-2s}}u)^{2^{2s}-1}e^{2^{-s}-2^s}d^{2^{2s}-1}.$$

Using the fact that 3 divides $2^r-1$ if and only if $r$ is even, we see that the right hand side of this expression is a cube while the left hand side is not. Thus, the kernel of $L_b$ has at most four elements.

\item[(iii)]{ We demonstrate that $e \neq 0$.}

For the sake of contradiction suppose that $e=0$. Then we have 
$$b^{2^{2k-s}+2^k-2^{2k}-2^{-k-s}}t^{2^{2k-s}-2^{-s}} = 1,$$

and hence

$$ (bt^{-1})^{2^{2k-s}+2^k-2^{2k}-2^{-k-s}}t^{2^{2k-s}-2^{-s}}=1.$$ 
Further rearrangement gives
\begin{equation}
 \ \ \ \ \ \ \ \ \ \ \ \ \ \ \ \ \ \ \ \ \ \ \ (b t^{-1})^{(1-2^k)(2^{2k-s}+2^k)} = t^{2^{-s}(1-2^{2k})}. \label{en0}
\end{equation}
As $4$ divides $k+s$, $2^{k+s} \equiv 1 \mod 5$.  Also $2^{2k}+1 \equiv 0 \mod 5$ for any odd $k$. Therefore $5$ divides $2^{k+s}+2^{2k}$ and hence $5$ divides $2^k + 2^{2k-s}$. The left hand side of (\ref{en0}) is a fifth power while the right hand side is not because $t$ is primitive and $2^{-s}(1-2^{2k})$ is not a multiple of $5$. We conclude that $e \neq 0$. 

\item[(iv)]{We next rule out the case $c=0$.}

Suppose $c=0$. Then we have

$$b^{2^{2k-s}+1-2^{-k-s}-2^{-k}} t^{2^{-k-s}+2^{-k}-2^{-s}-1} = 1,$$ 
from which we derive
$$ (bt^{-1})^{(2^k-1)(2^{-k}-2^{2k-s})} = t^{2^{-s}(2^{2k}-1)}. $$ 

By similar observations as before we can demonstrate that only the left hand side of the expression above is a fifth power. This gives us the desired contradiction and we conclude that $c \neq 0$. 

\item[(v)]{We show that if $d = 0$ then $L_b(u)=0$ has at most 4 solutions. }

Suppose that $d = 0$. Then Equation (\ref{4k3}) becomes 
\begin{equation}  \ \ \ \ \ \ \ \ \ \ \ \ \ \ \ \ \ \ \ \   c^{2^{2k}} u^{2^{-s}}+ e u^{2^{2k-s}}=0. \label{dn0}\end{equation}

Let $H(u):= c^{2^{-s}}u +e^{2^{-s}}u^{2^{2k}}$, so that solutions to (\ref{dn0}) comprise the kernel of $H(u)$. For some $v \neq 0$ in the kernel of $H(u)$, consider the equation 
$$H_v (u):=u H(u)+v H(v) + (u+v)H(u+v) = 0.$$
This yields   $H_v (u)=e^{2^{-s}}(u^{2^{2k}}v   +    v^{2^{2k}}u) = 0,$
from which we obtain $u^{2^{2k}}=v^{2^{2k-1}}u$. Applying this relation to $L_b(u)=0$ gives us the equation

$$L_b(u)=(b+(b  \alpha)^{2^k} v^{2^{2k+s}-2^s})u^{2^s} + (b^{2^{-s}}+(b \alpha)^{2^{-k-s}}v^{2^{2k-s}-2^{-s}})u^{2^{-s}}=0.$$
If both coefficients in the above expression are non-zero, then, by Corollary \ref{rsol}, it has at most four solutions. If exactly one of the coefficients is $0$, then $u=0$ is the unique solution. If both coefficients vanish, then we have
$$b^{2^{-s}}+(b \alpha)^{2^{k-s}} v^{2^{2k}-1}=0,$$
and
$$b + (b \alpha)^{2^{-k}}v^{2^{2k}-1}=0,$$
From which we derive 
$$v^{2^{2k}-1}=b^{2^{-k}-2^{k-s}}\alpha^{-2^{k-s}}=b^{1-2^{-k}}\alpha^{-2^{-k}}$$
which implies that $e = 0$, a previously established contradiction.


\end{enumerate}

This completes the proof of the theorem for the case when $f = f_1$. When $f=f_2$ a near identical proof applies. We simply interchange $k$ and $-k$ in all equations and use the fact that in this case $5$ divides $2^{k-s}-1$ to achieve the required contradictions concerning fifth powers.
\x

\section{Families (3) and (4)}

For proofs of the following theorems,
which compute the Fourier spectra of families (3) and (4),
 see \cite{BBMMcG2} and \cite{BBMMcG3} respectively.
We state the results here for completeness.

\begin{theorem}
Let $n=2k$ and let 
$$f(x) = \alpha x^{2^{s}+1}+{\alpha}^{2^k}x^{2^{k+s}+2^k}+\beta
x^{2^k+1}+\sum_{i=1}^{k-1} {\gamma}_i x^{2^{k+i}+2^i},$$
where $\alpha$ and $\beta$ are primitive elements of $L$, and
${\gamma}_i \in GF(2^k)$ for each $i$ and $(k,s)=1$. Then the Fourier spectrum of
$f(x)$ is $\{0, \pm 2^{\frac{n}{2}}, \pm 2^{\frac{n+2}{2}} \}.$
\end{theorem}

\begin{theorem}
Let $$f(x) = x^3 + Tr(x^9)$$ on $L$. Then the Fourier spectrum of $f(x)$ is $\{0, \pm 2^{\frac{n+1}{2}}\} $ when $n$ is odd and $\{0, \pm 2^{\frac{n}{2}}, \pm 2^{\frac{n+2}{2}} \}$
when $n$ is even.
\end{theorem}

\section{Closing remarks and open problems} For each of the above quadratic APN functions considered, the Fourier spectrum turned out to be the same as the Gold functions. The example of Dillon on $GF(2^6)$ cited in the introduction of this paper is the only known example of a quadratic APN function that does not have this spectrum. This means that the dual code of this function (as defined in Section 2) has the same minimum distance as the double error-correcting BCH code (the dual code corresponding to the function $x^3$), but has a different weight distribution.\\

{\bf Open problem 1}: Find other examples of quadratic APN functions for even $n$ 
that do not have the same Fourier spectrum as the Gold function $x^3$.\\

In \cite{BBMMcG} the following trinomial function (family (5) in the introduction)
over $GF(2^{3k})$ was shown to be APN:
$$f(x)= ux^{2^{-k}+2^{k+s}}+u^{2^{k}}x^{2^{s}+1}+vx^{2^{k+s}+2^{s}},$$
where $u$ is primitive, $v \in GF(2^k),  (s,3k)=1, (3,k)=1$ and 3 divides $k+s$.\\

{\bf Open problem 2}: Determine the Fourier spectrum of the above APN function.

\end{document}